# Deep learning-based segmentation of T1 and T2 cardiac MRI maps for automated disease detection

**Authors: Andreea Bianca Popescu**[1,2]**, Andreas Seitz**[3]**, Heiko Mahrholdt**[3]**, Jens Wetzl**[4]**, Athira Jacob**[5]**, Lucian Mihai Itu**[1,2]**, Constantin Suciu**[1,2]**, Teodora Chitiboi**[6]

**Affiliations:**

1. Department of Automation and Information Technology, Transilvania University of Brasov, 500024 Brasov, Romania,
2. Siemens SRL, 500097 Brasov, Romania,
3. Department of Cardiology and Angiology, Robert Bosch Hospital, 70376 Stuttgart, Germany,
4. Magnetic Resonance, Siemens Healthcare GmbH, 91052 Erlangen, Germany
5. Siemens Medical Solutions USA, Inc., Princeton, 08540 NJ, United States,
6. Siemens Healthineers AG, 20099 Hamburg, Germany

**Corresponding author:** Andreea Bianca Popescu

Department of Automation and Information Technology, Transilvania University of Brasov, 500024 Brasov, Romania, andreea.popescu@unitbv.ro

## Abstract

### Objectives

Parametric tissue mapping enables quantitative cardiac tissue characterization but is limited by inter-observer variability during manual delineation. Traditional approaches relying on average relaxation values and single cutoffs may oversimplify myocardial complexity. This study evaluates whether deep learning (DL) can achieve segmentation accuracy comparable to inter-observer variability, explores the utility of statistical features beyond mean T1/T2 values, and assesses whether machine learning (ML) combining multiple features enhances disease detection.

### Materials & Methods

T1 and T2 maps were manually segmented. The test subset was independently annotated by two observers, and inter-observer variability was assessed. A DL model was trained to segment left ventricle blood pool and myocardium. Average (A), lower quartile (LQ), median (M), and upper quartile (UQ) were computed for the myocardial pixels and employed in classification by applying cutoffs or in ML.

Dice similarity coefficient (DICE) and mean absolute percentage error evaluated segmentation performance. Bland-Altman plots assessed inter-user and model-observer agreement. Receiver operating characteristic analysis determined optimal cutoffs. Pearson correlation compared features from model and manual segmentations. F1-score, precision, and recall evaluated classification performance. Wilcoxon test assessed differences between classification methods, with $p < 0.05$ considered statistically significant.

### Results

144 subjects (mean age 42.2 years ± 16.1, 76 men) were split into training (100), validation (15) and evaluation (29) subsets. Segmentation model achieved a DICE of 85.4%, surpassing inter-observer agreement. Random forest applied to all features increased F1-score (92.7%, $p < 0.001$).

**Conclusion**

DL facilitates segmentation of T1/ T2 maps. Combining multiple features with ML improves disease detection.

**Key points**

*Question* Manual segmentation of myocardial T1/T2 maps is time-consuming and affected by inter-observer variability, while relying on single cutoffs values for diagnosis may oversimplify myocardial complexity.

*Findings* Deep learning achieves segmentation accuracy within inter-observer variability, while machine learning improves disease detection compared to singular cutoffs.

*Clinical relevance* Automated segmentation and feature extraction from T1/T2 maps can enhance workflow efficiency, reduce inter-observer variability, and improve diagnostic consistency. The high recall of the machine learning model minimizes missed diagnoses, ensuring more reliable disease detection.

**Abbreviations**

**A:** Average

**AUC:** Area under the curve

**DICE:** Dice similarity coefficient

**DL:** Deep learning

**FN:** False negatives

**FP:** False positives

**IoU:** Intersection over Union

**GT:** Ground truth

**KNN:** k-nearest neighbors

**LV:** Left ventricle

**LQ:** Lower quartile

**M:** Median

**MAPE:** Mean absolute percentage error

**ML:** Machine learning

**MM:** Model-generated Mask

**PCC:** Pearson correlation coefficients

**RF:** Random forest

**ROC:** Receiver operating characteristic

**SNR:** Signal-to-noise ratio

**SVM:** Support vector machines

**TN:** True negatives

**TP:** True positives

**UQ:** Upper quartile

**INTRODUCTION**

Cardiac MRI has become increasingly used in evaluating patients [1]. Parametric tissue mapping, which includes calculations of local T1 and T2 relaxation times, can facilitate quantitative cardiac tissue characterization [2]. However, the manual delineation of images is often labor-intensive, challenging for human experts, and highly subjective [3]. Establishing a consistent myocardium delineation protocol from T1 and T2 maps to improve results reproducibility is an active research topic [4, 5].

Deep learning (DL)-based segmentation has shown high potential in automating and standardizing cardiac MRI myocardium segmentation. Several studies have explored convolutional neural networks for T1 mapping segmentation, including applications in large patient cohorts [6], integration of quality control measures [7], and the use of synthetic contrast augmentation [8]. Combined segmentation of T1 and T2 maps [9, 10], as well as native and post-contrast T1 maps [11], has shown promising results in reducing the need for extensive annotated datasets. Additionally, transfer learning has proven effective in myocardium segmentation for T1/T2 maps [12, 13]. However, none of the above studies employed the segmented maps for disease detection.

Conventional classification of myocardial tissue abnormalities relies on average T1 and T2 values with predefined cutoff thresholds [14, 15, 16, 17]. This approach may oversimplify the T1 or T2 relaxation pattern, potentially overlooking nuanced tissue characteristics that could enhance diagnostic accuracy.

Recent research suggested that incorporating statistical features beyond mean values, such as quartile-based or radiomic features, may improve disease classification [18, 19]. Texture analysis applied to T1/T2 relaxation maps, combined with ML, improved liver fibrosis classification [20]. Still, the application of such analyses in myocardial tissue characterization remains underexplored.

In this study, we aim to develop and evaluate an automated approach for myocardial segmentation and disease classification using DL and ML. We assess whether DL can achieve segmentation accuracy comparable to human experts and investigate the added value of statistical features beyond mean T1/T2 values in improving disease classification. Additionally, we evaluate whether combining multiple statistical features with ML enhances diagnostic accuracy compared to single-threshold classification.

## MATERIALS AND METHODS

### *Data*

This was a retrospective analysis of anonymized clinical cardiac MRI data. Ethical approval was waived by the institutional review board.

Overall, 144 subjects (52 normal cardiac MRI, 49 myocarditis, 20 sarcoidosis, 23 systemic disease) were scanned on a 1.5-T MRI system (MAGNETOM Aera, Siemens Healthineers, Erlangen, Germany). Native and post-contrast T1 modified Look-Locker inversion recovery (MOLLI) and T2-prepared balanced steady-state free precession (bSSFP) maps were acquired (MyoMaps, Siemens Healthineers, Erlangen, Germany). The dataset included 1266 myocardial maps: 828 T1 maps and 438 T2 maps. The images were acquired in short-axis orientation, at a basal, mid-ventricular, or apical location, with an isotropic in-plane resolution in the range of $1.6x1.6 - 2.0x2.0$ mm$^2$, 8 mm slice thickness, flip angle 35° for T1 and 70° for T1, TE $0.97 - 1.09$ ms for T1 and $1.04 - 1.16$ for T2, TR $344 - 468$ for T1 and $104 - 285$ for T2. This dataset was denoted as Dataset$_A$. Demographic information and clinical characteristics are presented in Table 1. Representative examples from the "normal cardiac MRI", "sarcoidosis" and "myocarditis" groups are shown in Figure 1.

Table 1 Demographic information for the subjects in Dataset$_A$.

| Subject type | No. subjects (train/validation/test) | Age | Sex [No. males] | EF [%] | EDV [ml] | ESV [ml] |
|---|---|---|---|---|---|---|
| Total | 144 (100/15/29) | 42.2 ± 16.1 | 76 | 62.8 ± 9.4 | 137.8 ± 39.4 | 53.6 ± 28.5 |
| Normal cardiac MRI | 52 (36/6/10) | 39.5 ± 14.4 | 29 | 67.8 ± 6.4 | 129.2 ± 28.1 | 42.1 ± 13.7 |
| Myocarditis | 49 (34/5/10) | 39.2 ± 16.1 | 30 | 60.3 ± 9.3 | 146.9 ± 49.7 | 62.6 ± 37.4 |
| Sarcoidosis | 20 (14/2/4) | 54.6 ± 10.4 | 9 | 60.8 ± 10.4 | 139.5 ± 39.9 | 56.2 ± 25.5 |
| Systemic diseases | 23 (16/2/5) | 46.0 ± 18.9 | 8 | 57.9 ± 9.6 | 136.5 ± 33.4 | 59.2 ± 25.9 |

The "normal cardiac MRI" group included asymptomatic individuals with no known cardiovascular disease and a normal cardiac MRI according to expert reading (i.e. normal ejection fraction, normal left ventricular dimensions, no fibrosis or oedema according to T1/T2 mapping or LGE imaging). Most patients in this group underwent MRI for other reasons (e.g. dilated ascending aorta). The "myocarditis" group included patients with ≥1 clinical myocarditis criteria and ≥1 cardiac MRI abnormality (myocardial oedema or LGE with myocarditis pattern) [21]. Clinical criteria involved symptoms such as acute chest pain, dyspnea, palpitations, arrhythmia, fatigue, and syncope. Patients with history of coronary artery disease, myocardial infarction, or previous revascularization were excluded. Individuals included in the sarcoidosis and systemic disease were included irrespective of proof of myocardial involvement, as long as their systemic disease was confirmed. Proving myocardial involvement in these diseases is challenging (e.g. sampling error, previous vs. current involvement, etc.).

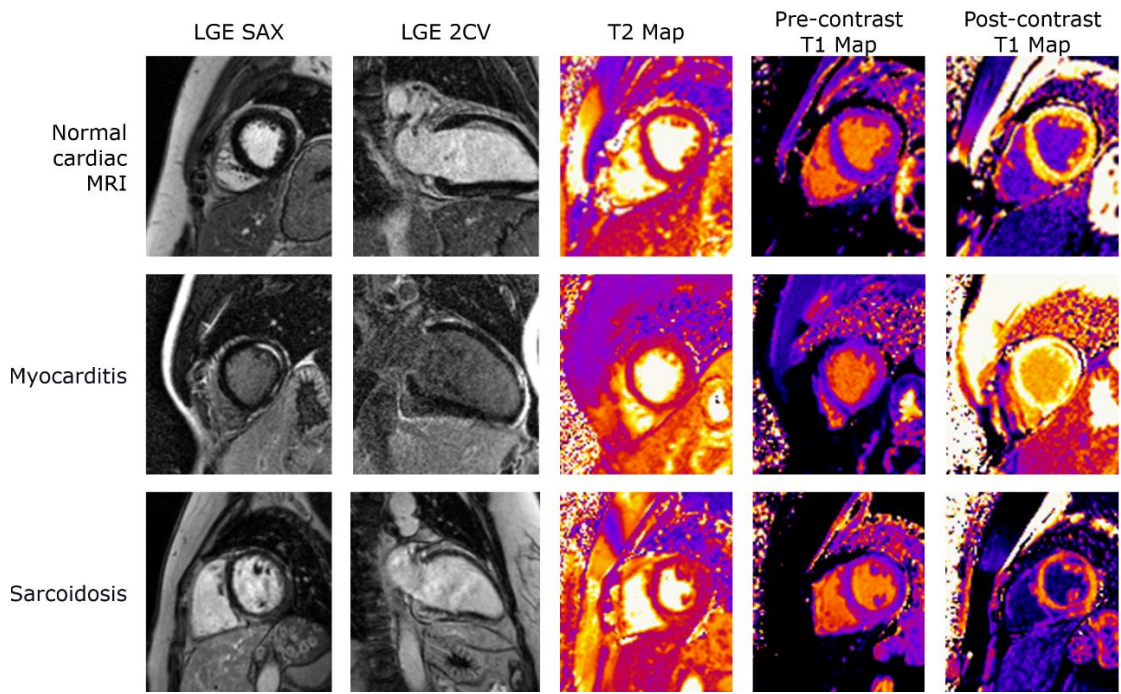

Figure 1 Example images illustrating both healthy and diseased cases across multiple imaging modalities. For one patient in the "normal cardiac MRI", "myocarditis" and "sarcoidosis" groups, five images are shown: late gadolinium enhancement (LGE) short-axis (SAX), LGE 2-chamber view (2CV), T2 map, pre-contrast T1 map, and post-contrast T1 map.

Greulich et al. [16] demonstrated abnormal myocardial T1 and T2 mapping values in patients with systemic sarcoidosis compared to healthy individuals independent of the presence of LGE. Although certain LGE patterns are indicative of cardiac sarcoidosis, there are currently no validated universal criteria for cardiac involvement of systemic sarcoidosis or isolated cardiac sarcoidosis, respectively. In the present study, patients were classified as having sarcoidosis when the following criteria applied [16]: (a) systemic sarcoidosis diagnosed by biopsy or clinical criteria, and (b) no history of coronary artery disease, myocardial infarction, or previous revascularization. Similar to sarcoidosis, previous studies have demonstrated abnormal myocardial T1 and T2 values in patients with systemic inflammatory and autoimmune diseases, suggesting myocardial involvement (22, 23). The "systemic disease" group included patients with collagenoses (e.g., systemic lupus erythematosus, systemic sclerosis), vasculitis (e.g., eosinophilic granulomatosis with polyangiitis), IgG4-related disease, muscular dystrophy, and rheumatoid arthritis.

The subjects were divided into subsets of 100 for training, 15 for validation, and 29 for testing. Details on data distribution among subsets are included in Table 2.

Table 2 Summary of datasets and distribution of data among training, validation, and testing subsets.

|  | Data | Total | Train | Validation | Test |
|---|---|---|---|---|---|
|  | Patients | 144 | 100 | 15 | 29 |
|  | T1 maps | 828 | 576 | 81 | 171 |
|  | T1 pre | 412 | 284 (99 patients) | 42 | 86 |
| Dataset$_A$ | T1 post | 416 | 292 (98 patients) | 39 (14 patients) | 85 |
|  | T2 maps | 438 | 301 | 47 | 90 |
|  | Total images | 1266 | 877 | 128 | 261 |
| Dataset$_B$ (pretraining) | Patients | 192 | 183 | 9 |  |
|  | T1-weigthed Images | 10560 | 10065 | 495 | - |

In the T1 and T2 maps, the myocardium was manually segmented by two observers with 3 (Observer 1) and 5 (Observer 2) years of experience in annotating cardiac anatomy. They used an internally-developed annotation tool based on ProjectX (https://github.com/proyecto26/projectx) to draw the endocardial and epicardial contours with instructions to exclude papillary muscles, trabeculations, and pixels affected by partial volume effects. The annotations were supervised and validated by a radiologist with 11 years of cardiac MRI experience. Training and validation data were randomly assigned to one of the two observers. The entire test subset was independently annotated by both observers, in random order and without time delays between samples. The inter-observer variability in delineating the left ventricle (LV) blood pool and myocardium was assessed on the test subset using dice similarity coefficient (DICE) and mean absolute percentage error (MAPE). DICE was calculated to measure the spatial overlap of the segmentations, where higher values indicate better agreement. MAPE was used to quantify the percentage difference in the T1/T2 values extracted from the myocardium.

Abnormal cardiac MRI datasets with inflammatory or infiltrative myocardial disease (i.e., myocarditis, sarcoidosis, systemic diseases) were grouped into a single category labeled "diseased," allowing disease detection to be treated as a binary classification problem.

An additional 192 subjects (Dataset$_B$) were used for pre-training the segmentation network. This dataset included a total of 10,065 T1-weighted images with publicly available annotations [22] and was acquired on a 1.5-T MRI system (Achieva, Philips Healthcare, Best, Netherlands). Details on the data split are provided in Table 2. The pre-training process involved initializing the segmentation network using Dataset$_B$ to learn general features related to T1-weighted MRI segmentation, improving its ability to generalize to new data. These pre-trained weights were then fine-tuned on Dataset$_A$, enabling the network to specialize in segmenting the specific T1/T2 maps used in this study.

### *Mapping Segmentation*

A DenseUnet [23, 24] model was trained for LV blood pool and myocardium segmentation in T1 (pre- and post-contrast images) and T2 maps. The segmentation masks contained three class labels: LV blood pool (everything inside the endocardial contour), myocardium (between the endocardial and the epicardial contour), and background (everything outside the epicardial contour). The architecture comprised five pooling layers with convolutions of 3x3.

The Jaccard loss was calculated according to Eq. 1, using the Intersection over Union (IoU) between the ground truth (GT) and the model-generated mask (MM). The loss function was minimized using the Adam optimizer. The learning rate and the batch size were empirically set to $3 \cdot 10^{-4}$ and 2, respectively. First, pretraining was performed on $Dataset_B$ for 100 epochs. Then, the model was finetuned on $Dataset_A$ for another 100 epochs. Dropout was employed to prevent overfitting by removing 20% of the connections.

$$Loss = 1 - IoU = 1 - \frac{GT \cap MM}{GT \cup MM} \qquad \text{(Eq. 1)}$$

All images used for training were resampled to 1x1 mm resolution and normalized using the $1^{st}$ and $99^{th}$ percentiles to scale pixel values to [0, 1]. The images were cropped to 288x288 mm around the image center. Intensity augmentations, including contrast stretch and Gaussian noise addition, as well as geometric augmentations such as random rotation, translation and vertical or horizontal flip, were applied during training.

*Feature Analysis and Disease Detection*

Once the myocardium mask was automatically extracted for the pre-contrast T1 and T2 maps, statistical features of the myocardium pixel were computed (A, LQ, M, and UQ). For each patient, the statistical features were averaged over all slices acquired separately for native T1 and T2 maps.

Receiver operating characteristic (ROC) curves were generated using the combined training and validation data to guide threshold selection. Similar to [25, 26, 27], optimal cutoff values were determined by maximizing Youden's J statistic (Eq. 2), which takes into account true positives (TP), false positives (FP), true negatives (TN), and false negatives (FN). These cutoff values were then used as classification criteria for the test subset, where values exceeding the threshold indicated a diseased patient.

$$J = \frac{TP}{TP+FN} + \frac{TN}{TN+FP} - 1 \qquad\qquad\qquad \text{(Eq. 2)}$$

Several ML classifiers were trained using multiple statistical features as input. Five classifiers were evaluated: logistic regression, k-nearest neighbors (KNN), support vector machines (SVM), random forest (RF), and Perceptron (a single artificial neuron), each trained using various combinations of T1 and T2 features on the training data subset. In one instance, feature selection was based on the ROC analysis, choosing the top two best-performing features for native T1 and T2 maps. In another set of experiments, all features from native T1 and T2 maps were used. Optimal hyperparameters were selected via grid search, maximizing the F1-score on the validation subset.

All experiments were run on a Tesla V100 SXM2 GPU (NVIDIA, Santa Clara, CA, USA) with 16 GB of dedicated memory. ML models were developed using PyTorch (version 1.12.1, https://pytorch.org/) and scikit-learn (version 1.2.1, https://scikit-learn.org/stable/).

***Statistical Analysis***

Segmentation performance was evaluated using DICE and MAPE and compared to the GT annotations provided by the two readers. MAPE was calculated according to Eq. 3, where G and M denoted the average T1 or T2 values from the GTs and MMs, respectively. This metric provided valuable insights into the impact of segmentation accuracy on relaxation time calculation.

$$MAPE = \frac{G-M}{G} \cdot 100 \qquad\qquad\qquad \text{(Eq. 3)}$$

The impact of the segmentation method (manual or automatic) on statistical feature extraction was evaluated by computing Pearson correlation coefficients (PCC). Bland-Altman plots were created to assess the inter-user agreement and the agreement between the model and the observers for the statistical features of the T1/T2 values.

ROC analyses conducted on the combined training and validation subsets assessed the discriminative power of each feature. Additionally, the area under the curves (AUCs) and the corresponding confidence intervals were calculated to quantify the overall performance of each feature. AUCs were compared for statistical significance using the method proposed by DeLong et al. [28].

The classification performance on the test subset was evaluated using precision, recall and F1-score. Differences between the classification approaches were tested using Wilcoxon test. A $p$-value below 0.05 was considered statistically significant.

**RESULTS**

*Mapping Segmentation*

A comparison between automatic segmentation masks and annotations from two observers are presented in Table 3.

Table 3 Comparisons between the model-generated masks, the annotations provided by Observer 1, the annotations provided by Observer 2.

| Data | Nr. of images | LV DICE [%] | | | MYO DICE [%] | | | MYO MAPE [%] |
|------|------|------|------|------|------|------|------|------|
| | | Mean | Median | Minimum | Mean | Median | Minimum | |
| Model vs Observer 1 | | | | | | | | |
| All | 261 | 96.8 | 97.4 | 88.1 | 87.0 | 88.4 | 60.7 | 1.6 |
| T2 | 90 | 96.2 | 96.9 | 88.1 | 88.9 | 90.3 | 73.6 | 2.7 |
| T1 Pre | 86 | 97.3 | 97.6 | 91.1 | 87.1 | 88.5 | 73.0 | 1.2 |
| T1 Post | 85 | 97.0 | 97.4 | 88.9 | 84.8 | 86.4 | 60.7 | 0.8 |
| Model vs Observer 2 | | | | | | | | |
| All | 261 | 96.0 | 96.7 | 81.2 | 83.8 | 86.1 | 39.0 | 1.9 |
| T2 | 90 | 95.5 | 96.6 | 76.2 | 86.2 | 87.2 | 65.2 | 3.5 |
| T1 Pre | 86 | 96.5 | 97.2 | 94.9 | 83.6 | 86.4 | 64.3 | 1.3 |
| T1 Post | 85 | 96.0 | 96.7 | 81.2 | 81.3 | 83.6 | 39.0 | 0.9 |
| Observer 1 vs Observer 2 | | | | | | | | |
| All | 261 | 95.4 | 96.3 | 77.3 | 81.6 | 83.7 | 47.7 | 2.5 |
| T2 | 90 | 94.8 | 96.1 | 79.7 | 85.4 | 85.4 | 69.7 | 4.7 |
| T1 Pre | 86 | 96.0 | 96.7 | 84.1 | 81.5 | 83.7 | 51.8 | 1.5 |
| T1 Post | 85 | 95.3 | 96.2 | 77.3 | 77.5 | 77.9 | 47.7 | 1.1 |

The model achieved an average DICE of 85.4% $\pm$ 1.6% and a MAPE of 1.75% $\pm$ 0.15%. While the DICE for the T2 maps surpassed that of T1, the lower MAPE in T1 implied that T1 values were less affected by segmentation variations than T2. Additionally, Table 3 shows the inter-observer variability in delineating the LV blood pool and the myocardium. The level of agreement between the model and each observer was higher than that between the two observers, which yielded a DICE of 81.6% and a MAPE of 2.5%. Figure 2 shows two examples with low inter-observer agreement where the automatic contours closely match the annotations of Observer 1.

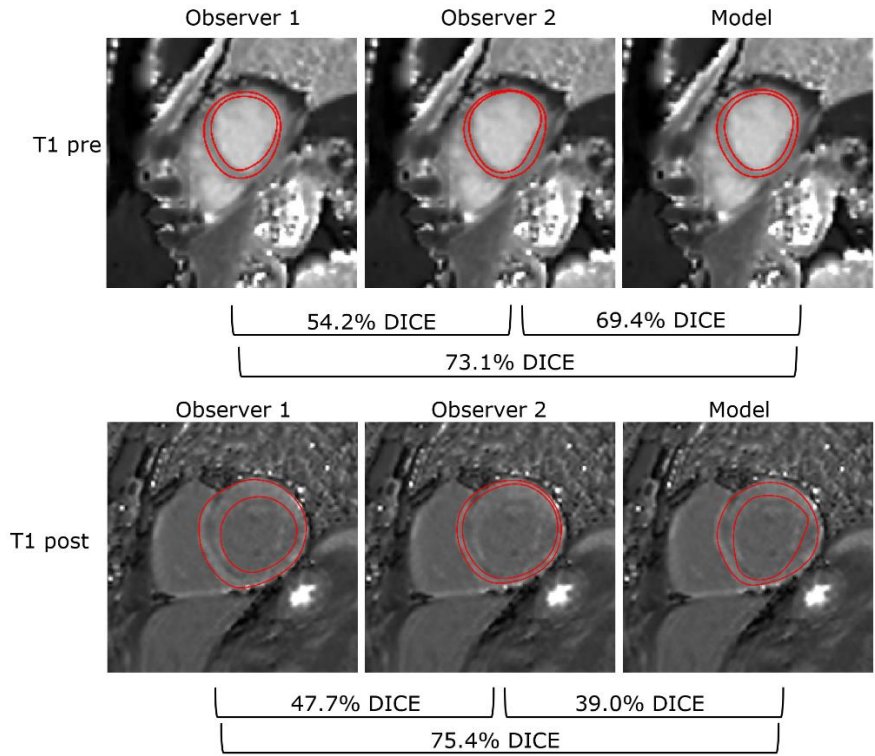

Figure 2 Examples of disagreement between observers in segmenting a pre-contrast and a post-contrast T1 map.

### Feature Analysis and Disease Detection

Table 4 shows the PCC computed between features derived from automatic and manual segmentations. The model results showed strong agreement (PCC > 0.9) with expert annotations, especially for T1 mapping and specific T2 mapping features (LQ and M), and all correlations were statistically significant ($p < 0.001$), confirming that small segmentation variations had minimal impact on the T1/T2 quantification. Similar to MAPE, PCC indicated that segmentation variations have a greater impact the statistical features obtained from T2.

Table 4 Pearson correlation coefficients computed for the features derived based on the automatic contours (auto) and based on the contours provided by the two observers.

| Experiment | $\text{PCC}_{\text{obs1/obs2}}$ | $\text{PCC}_{\text{obs1/pred}}$ | $\text{PCC}_{\text{obs2/model}}$ |
|---|---|---|---|
| T1, A | 0.961 | 0.984 | 0.976 |
| T1, LQ | 0.960 | 0.992 | 0.975 |
| T1, M | 0.976 | 0.992 | 0.985 |
| T1, UQ | 0.960 | 0.977 | 0.976 |
| T2, A | 0.886 | 0.929 | 0.960 |
| T2, LQ | 0.991 | 0.994 | 0.996 |
| T2, M | 0.975 | 0.983 | 0.988 |
| T2, UQ | 0.874 | 0.920 | 0.950 |

The Bland-Altman plots shown in Supplementary Figures 1 and 2 illustrate inter-user agreement and the agreement between the model and the observers for the statistical features. The feature values obtained from automatic and manual masks were relatively consistent. The results showed narrow limits of agreement and small mean biases.

The ROC curves and AUC scores in Figure 3 show the classification power of the automatically extracted statistical features in distinguishing between healthy and diseased subjects. All T1 features achieved AUC scores above 70%. The T1 map features with the highest AUC were T1 UQ (AUC = 75.2%), followed by T1 A (AUC = 73.8%). For T2, the best performers were T2 UQ (AUC = 63.5%) and T1 A (AUC = 62.6%). The DeLong test showed that the only statistically significant differences were between the T1 LQ and T2 LQ ($p = 0.047$), as well as between T1 M and T2 M ($p = 0.03$).

Table 5 (top) contains the classification results obtained on the held-out, test subset for the optimal cutoff values derived for each feature. The best performance was achieved by T1 LQ and UQ, both achieving an F1-score of 66.7% and a precision of 90.1%. T2 features showed lower classification performance. The only significant difference was observed between T1 LQ and T2 LQ ($p = 0.03$). Recall values were consistently lower than precision.

Table 5 Classification performance: achieved on the test subsets when applying the optimal cutoff value derived from the training and validation subsets by maximizing Youden's J statistic (top) and achieved on the test subsets by each machine learning classifier (bottom).

| Approach | Feature(s) | Optimal cutoff value [ms] | F1-score [%] | Precision [%] | Recall [%] |
|---|---|---|---|---|---|
| Cutoff | T1 A | 989 | 62.1 | 90.0 | 47.4 |
| | T1 LQ | 942 | 66.7 | 90.1 | 52.6 |
| | T1 M | 988 | 62.1 | 90.0 | 47.4 |
| | T1 UQ | 1034 | 66.7 | 90.1 | 52.6 |
| | T2 A | 54 | 33.3 | 80.0 | 21.5 |
| | T2 LQ | 49 | 33.3 | 80.0 | 21.5 |
| | T2 M | 52 | 33.3 | 80.0 | 21.5 |
| | T2 UQ | 57 | 46.2 | 85.7 | 31.6 |
| Logistic regression | [T1 LQ, UQ] or [T1 A, LQ, M] | - | 90.0 | 85.7 | 94.7 |
| KNN | [T1, A, UQ] | - | 85.0 | 81.0 | 89.5 |
| SVM | [T1, A, UQ] or [T1, A, LQ, M, UQ] | - | 87.2 | 85.0 | 89.5 |
| Random forest | [T1, A, LQ, M, UQ, T2, A, LQ, M, UQ] | - | 92.7 | 86.4 | 100 |
| Perceptron | [T1, A, UQ] or [T1, A, LQ, M, UQ] | - | 92.3 | 90.0 | 94.7 |

Table 5 (bottom) presents the best classification performance on the test subset for each type of ML classifier evaluated in this study. The KNN model achieved the highest performance using T1 A and UQ as inputs, with an F1-score of 85.0%, significantly surpassing the threshold-based method on T1 UQ. Training an SVM model on the same features resulted in an F1-score of 87.2%, though the improvement over KNN was not statistically significant ($p = 1$). Combining T1 LQ and UQ, or T1 A, LQ, and M, resulted in the best performance for the logistic regression model, achieving an F1-score of 90%. This performance was not significantly different from KNN or SVM. Similarly, the Perceptron achieved an F1-score of 92.3% using the same feature set as SVM, but without a significant improvement over the other algorithms. Among the RF experiments, combining all features produced the highest F1-score (92.7%) and a recall of 100%. This performance was significantly higher than that of the threshold-based method on T1 UQ ($p < 0.001$), but not statistically significant compared to other ML classifiers ($p = 0.57$).

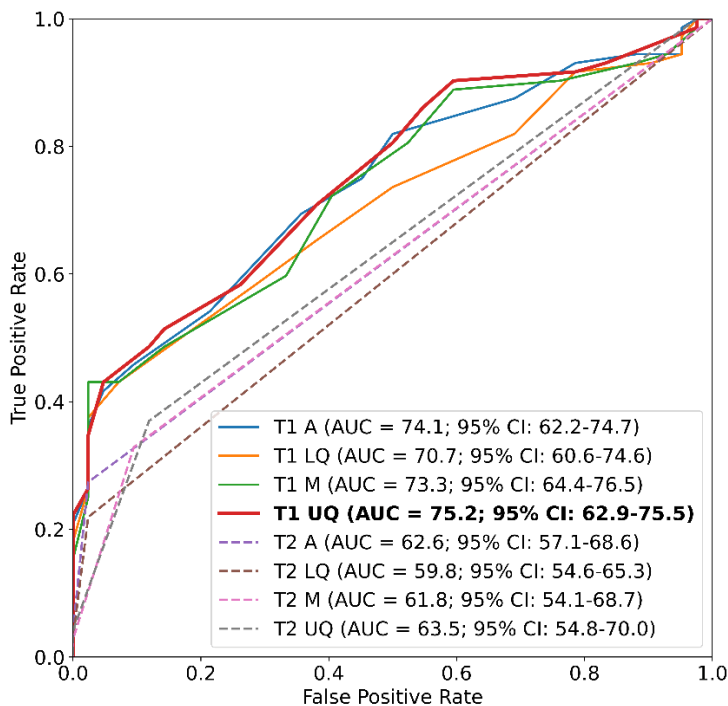

Figure 3 The ROC curves, AUC scores and the corresponding confidence intervals (CI) for all features evaluated as classification criteria. AUC values are expressed as percentages [%]. Statistical features such as the average (A), lower quartile (LQ), median (M) and upper quartile (UQ) were derived based on the automatic myocardium masks for the T1 and T2 maps from the train and validation subsets combined.

The average processing time per patient was $0.4 \pm 0.03$ seconds, with most time spent on myocardium segmentation. Feature extraction required ~0.1 seconds, while the classification itself took under one millisecond.

## DISCUSSION

This study explored multiple disease detection approaches using T1/T2 mapping MRI. The segmentation model showed higher agreement with each observer than the observers had with each other. Among the analyzed features, UQ was the strongest disease indicator in both T1 and T2 maps, though not significantly better than A. Moreover, using individual features to perform a simple threshold-based classification yielded unsatisfactory results. Single-feature, cutoff-based classification performed poorly, whereas combining multiple features in a ML classifier significantly improved results. Classifier choice had minimal impact on performance.

The proposed approach has the potential to augment the manual analysis of myocardial T1/T2 maps for the assessment of inflammatory or infiltrative myocardial diseases by automating segmentation and feature extraction – tasks which are often labor-intensive and prone to variability. The method's agreement with manual annotations suggests it could contribute to more consistent diagnoses.

The high recall of the RF model, especially when using all four features, resulted in no FN, reducing the risk of missed diagnoses. However, this high recall may increase FP, increasing the risk of over-diagnosis. Balancing recall and precision remains essential for clinical application. Future work could explore strategies to reduce FPs while maintaining high recall.

The segmentation model was based on a U-Net architecture with embedded dense blocks, similar to prior studies [29, 30]. While DL has been widely used for cardiac MRI segmentation [6, 7, 8, 9, 10, 11, 12, 13], few studies integrate both segmentation and disease detection [31, 32, 33]. Isensee et al. developed a fully automated pipeline for cardiac cine MRI [31]. Their approach utilized an ensemble of U-Net-inspired architectures for segmentation and classified patients into four pathology groups and one healthy group using a multi-layer perceptron and RF. While not directly comparable, their classification performance (accuracy of 92–93%) aligns closely with that achieved in this study (F1-score of 92.7%). Their system required only a few seconds per case, while our pipeline runs in under one second.

In this study, the ML classifiers were selected for their efficiency, diverse classification approaches, interpretability, and prior success in medical imaging [37, 38, 39]. Future studies could explore the benefits of more complex algorithms for this application.

This study used post-contrast T1 data when training the segmentation network to improve generalizability by exposing the model diverse imaging characteristics. However, for disease detection, only native (pre-contrast) T1 data were used, as not all patients may undergo post-contrast imaging in practice. Future studies could investigate the diagnostic potential of statistical features derived from post-contrast T1 maps.

Classification experiments revealed that T2 features consistently underperformed compared to T1, likely due to several factors. Higher inherent noise in T2 maps may reduce pixel intensity reliability, impacting the quality of the extracted statistical features. Although T2 segmentation performance in this study was higher than for T1, slight inaccuracies in myocardium delineation could still influence T2 features computation.

This study included a wide range of diseases without distinguishing between subsets where T1 and T2 mapping might be more sensitive, such as acute versus chronic conditions. Future research could explore larger cohorts to assess differences in classification accuracy between these subgroups. Nevertheless, investigating whether T1 and T2 mapping can differentiate normal from abnormal myocardium in a mixed, consecutive clinical patient cohort remains valuable.

Limitations include single-center data and the use of a single scanner, which may affect generalizability, as T1/T2 values vary across institutions and protocols [40]. Unless such variations can be mitigated in the acquisition protocol or post-hoc during data analysis [41], the automatic classifier for would need re-calibration for each new center. Additionally, the limited number of patients could impact the robustness of the results, potentially limiting the detection of subtle differences in statistical features and the generalizability of the classifiers. Larger, more diverse cohorts are needed to validate and enhance the reliability of the model. Also, more comprehensive diagnostic information beyond T1 and T2 mapping data could be required to differentiate between various disease classes effectively.

In conclusion, this study demonstrated the feasibility of automated myocardial segmentation in T1/T2 maps using DL and showed that combining multiple statistical features with ML improves disease detection in inflammatory/infiltrative cardiac conditions.

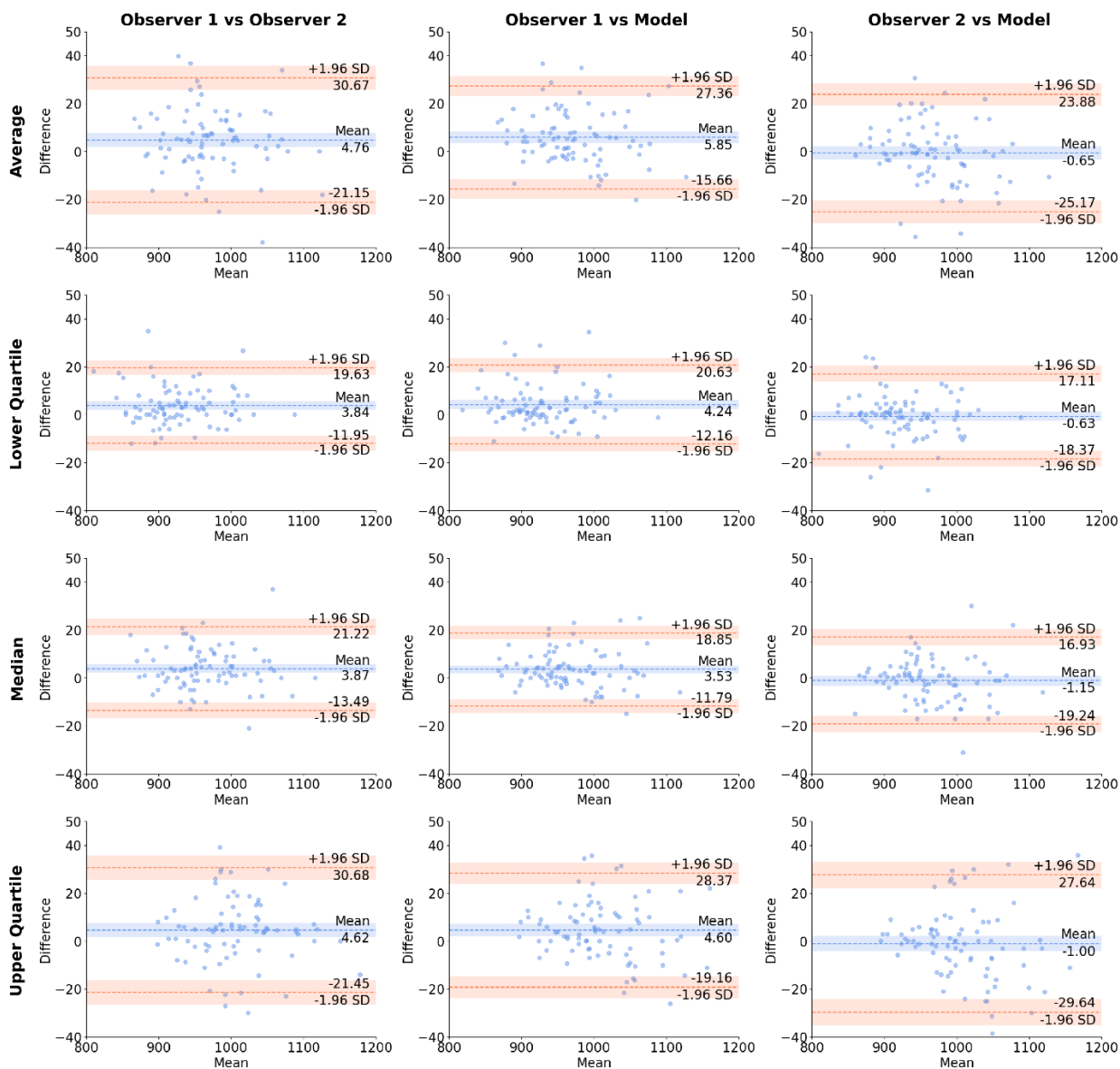

Supplemental Figure 1. Bland-Altman graphs illustrating the agreement between different users and the model in relation to various statistical features calculated from the segmentation of the T1 maps. The plots demonstrate relatively consistent feature values between automatic and manual segmentations, with narrow limits of agreement and small mean

biases. The agreement between the model and Observer 2 is higher, while the agreement levels for Observer 1 vs Observer 2 and Observer 1 vs the model are comparable.

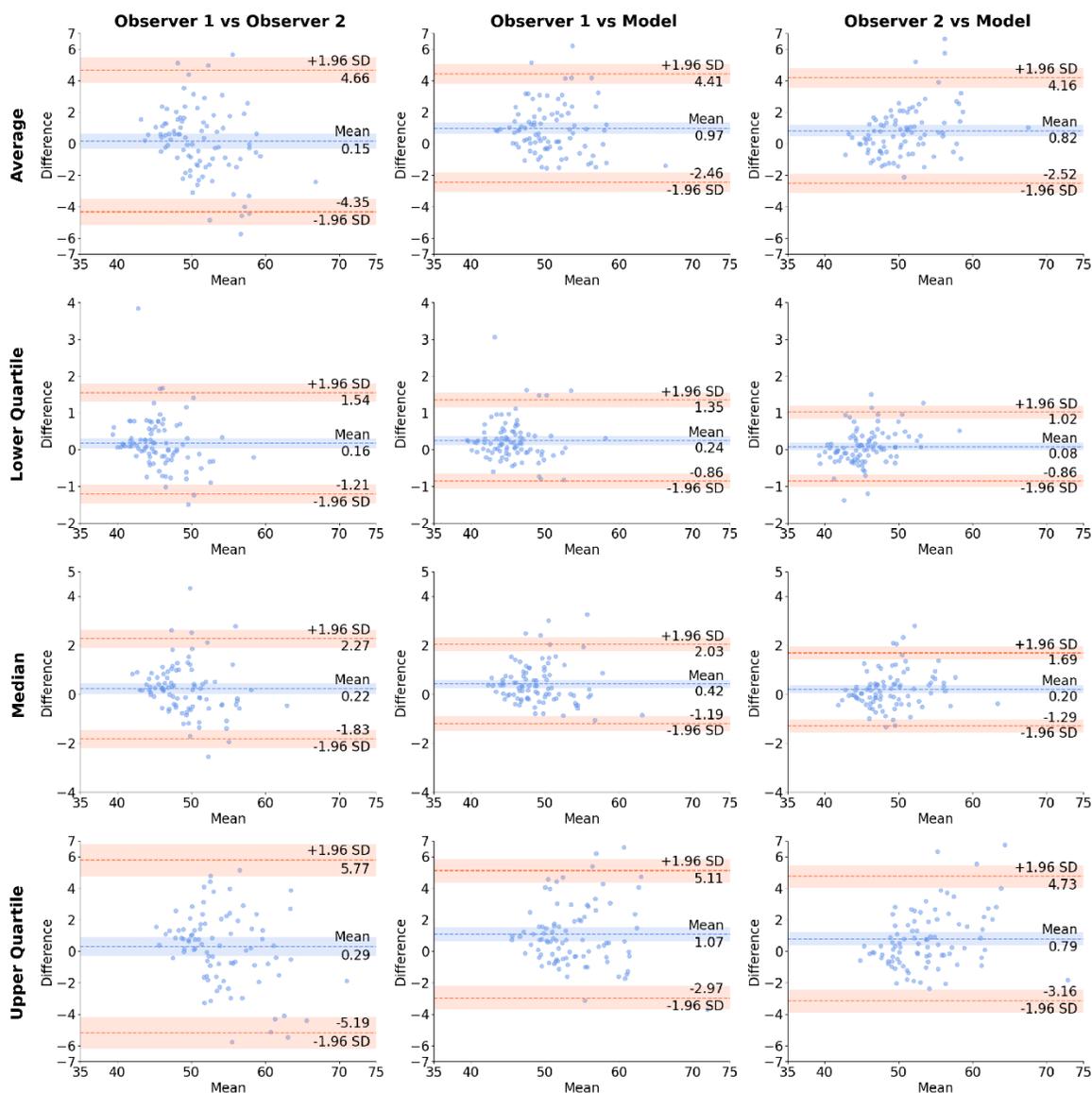

Supplemental Figure 2. Bland-Altman graphs illustrating the agreement between different users and the model in relation to various statistical features calculated from the segmentation of the T2 maps. The plots show very small mean biases that are comparable across all three analyses. However, as a general trend, the inter-user agreement appears slightly higher than the agreements between the model and the observers.